\pdfoutput=1
\documentclass[11pt]{article}

\usepackage[margin=1in]{geometry}
\usepackage{amsmath}
\usepackage{amssymb}
\usepackage{newtxtext,newtxmath}
\usepackage[numbers,sort&compress]{natbib}
\usepackage[utf8]{inputenc}
\usepackage[T1]{fontenc}
\usepackage{graphicx}
\usepackage{booktabs}
\usepackage{array}
\usepackage{tabularx}
\usepackage{subcaption}
\usepackage{siunitx}
\usepackage[hyphens]{url}
\usepackage[hidelinks]{hyperref}
\newcolumntype{Y}{>{\raggedright\arraybackslash}X}

\begin{document}

\title{When appearance and motion disagree: evaluating class--motion consistency against disguised uncrewed surface vessels}

\author{Jon Reifschneider\\Duke University, Durham, NC, USA\\\texttt{jon.reifschneider@duke.edu}}
\date{}
\maketitle

\begin{abstract}
Defense systems for maritime assets must distinguish approaching benign craft from uncrewed surface vessels (USVs) with sufficient warning to act. Shore cameras using computer vision can separate hostile vessels from benign traffic, but these systems may be attacked using an adversarial hull patch or physical disguise. We evaluate adversarial hull patches and find they suppress detection in up to 29\% of attempts but never cause misclassification as benign. Physical disguise, however, can bypass both a camera check and a radar presence check. To evaluate defenses, we develop a hybrid replay protocol from automatic identification system (AIS) tracks that combines recorded benign motion with direct or off-axis attack run-ins. We find that for recreational-labeled vessels attacking via direct run-in within 4 nautical miles, an asset-relative closing-rate check flags 94--95\% at 15--20 knots, compared with 12\% for a track-only speed check. The checks are complementary on off-axis approaches, constraining the attacker to remain below both thresholds and extending the time required to reach the asset. Flagging on either check has a 4.9\% held-out false-alarm rate overall and 12.3\% on recreational traffic. The replay-based evaluation protocol supports comparative test and evaluation of defense methods.
\end{abstract}


%% main text
\section{Introduction}\label{sec:intro}

Uncrewed surface vessels (USVs) have become increasingly prominent in recent maritime conflicts. Ukrainian craft have struck warships \cite{raveendran2025,wertheim2025} and port infrastructure \cite{raveendran2025} across the Black Sea, and Houthi craft have struck merchant shipping in the Red Sea \cite{ambrey2024,cardash2024}. Documented Houthi boats include a modified patrol boat, a purpose-built hull \cite{cardash2024}, and fishing boats used as camouflage in confirmed attacks \cite{ambrey2024,cardash2024}. Such a craft can inflict significant damage on a shoreline asset, and the attack develops within minutes, leaving the defender little time to respond. Defense analysts accordingly call for dedicated medium-range and close-in countermeasures against USVs \cite{boretti2026}.

What makes a USV particularly difficult to stop is that its appearance can be manipulated to mimic a legitimate vessel. Boats in these attacks have carried dummies, which can disguise them as crewed fishing craft \cite{ambrey2024,cardash2024}. A USV can alternatively be configured to pass as a sailing yacht or a recreational speedboat. The approaches to a working port often include large numbers of benign small craft, complicating the challenge of identifying a hostile vessel surrounded by benign traffic. Methods to distinguish a hostile craft must utilize either its appearance or its motion, both of which are controlled by the attacker.

Defense at a fixed asset is typically layered across sensors, patrol craft, and weapons. The sensor layer must identify which approaching boat poses a potential threat and how quickly a response is needed. No single channel can resolve that question for a small craft. An automatic identification system (AIS) report is self-generated and easily withheld. A radar track provides insight on motion but not the vessel type, and human operators can miss salient events during sustained monitoring (Section~\ref{sec:identity-channels}). Electro-optical and infrared cameras with machine-learning classifiers, now fielded commercially for this purpose \cite{seaai2025,anduril_sentry}, assign vessel classes to approaching craft at a standoff distance large enough to be useful. Camera systems in other safety-critical use cases, however, have been successfully attacked: stickers on a stop sign have made a traffic-sign classifier read a speed limit, and patches printed on the road surface have steered a production lane-centering system out of its lane \cite{eykholt2018,drp2021,pa_survey2024}.

\begin{figure}[tbp]
  \centering
    \includegraphics[width=\linewidth]{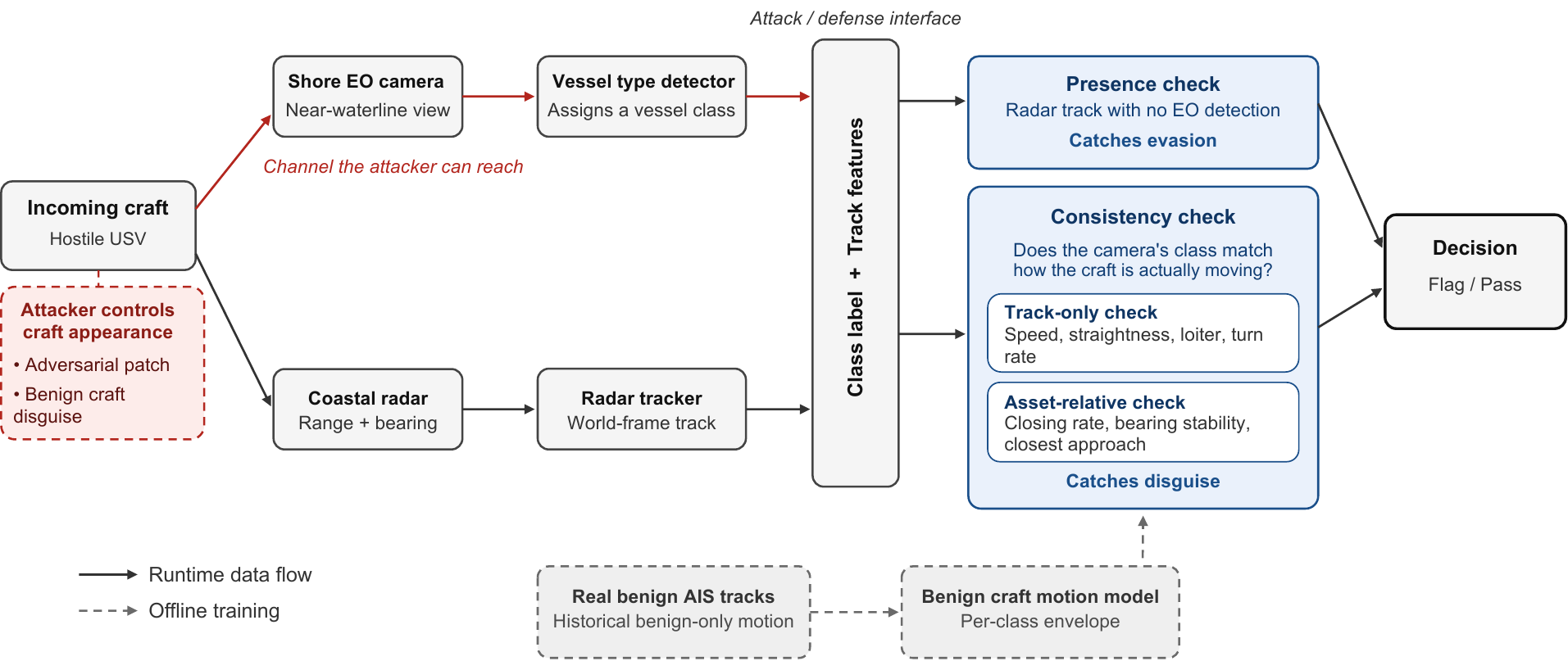}
    \caption{The attacker can change the craft's appearance and choose its physical motion but cannot falsify the radar measurement. Solid arrows are runtime data flows and the dashed path at the bottom is the benign vessel motion model, fit offline on historical vessel tracks.}\label{fig:system}
\end{figure}

A shore-based camera for port defense can provide a vessel class, but it relies on the craft's appearance, which can be manipulated by an attacker. Radar can measure motion that an attacker cannot easily falsify, but it does not identify the type of vessel. These limitations motivate a cross-modal consistency check: compare the vessel class assigned by the camera with the radar-measured track and determine whether a vessel of that class is likely to move that way. Figure~\ref{fig:system} shows the intersection of the two sensing channels.

Evaluating defense methods against this threat model is difficult because recorded attacks are scarce, field trials are hazardous, and actual benign traffic is needed to measure alert burden. An evaluation protocol was therefore developed that combines recorded approach histories with controlled hostile maneuvers. The factors varied across scenarios are exposed, and detection is reported alongside false alarms and warning times. Defense modeling studies likewise use designed simulation experiments to compare alternatives that cannot be tested safely in live operations, such as varying drone autonomy against an integrated air defense system \cite{martinez2026}. In the maritime domain, agent-based simulation of port defenses has shown that a small-vessel attacker becomes harder to intercept as benign traffic increases \cite{harris2013}.

We use the hybrid replay protocol to evaluate attacks against the camera layer and a cross-modal defense, with three findings. First, the tested physical patches suppressed detection in a minority of attempts, less often as the attacker's access to the model was reduced, but none of the 420 tested configurations produced a hostile-to-benign label change (Section~\ref{sec:rq1}). Physical disguise needs no model access and can produce the benign appearance assumed in the defense evaluation. A presence check, which alarms on a radar track without a camera detection, does not cover that case. We therefore evaluate the defense conditional on a benign camera label (Section~\ref{sec:threat}).

Second, we compare a track-only speed-cutoff check with an asset-relative closing-rate check on tracks that initially replay a recorded approach from a benign vessel before a simulated attack run-in. AIS reports are used to derive the tracks as a stand-in for radar measurements. On direct 15--20\,kn run-ins, the closing-rate check flags 94--95\% while the speed-cutoff check flags 12\% (Section~\ref{sec:speeds}). 

Third, the two checks cover different motion strategies. The speed check exposes a fast approach engineered for low closing rate. Motion below both cutoffs is not detected, but it imposes a cost on the attacker by taking longer to reach the asset. From 6\,nm the closing-rate cutoff implies a minimum arrival time of about 31 minutes. Warning time from the flag, however, can be much shorter: a track with a late turn toward the asset can reduce warning to about a minute (Section~\ref{sec:offset}). Flagging on either check has a 4.9\% held-out false-alarm rate overall and 12.3\% on recreational traffic. Class conditioning the checks improves their detection rate as compared to using common checks across all vessel types (Section~\ref{sec:class-value}). On direct run-ins the simple checks prove more effective than model-based comparison scorers (Table~\ref{tab:scorers}).

This paper makes three contributions: a class--motion defense, an evaluation protocol for testing that defense, and a maritime camera-attack evaluation that distinguishes failures by their consequences in a layered defense system.

\begin{enumerate}
\item We develop a class--motion consistency defense that compares a camera-assigned vessel class with independently measured speed and asset-relative closing rate. These simple checks cover complementary approach geometries and detect more than fitted model-based scoring methods (Sections~\ref{sec:scorer}, \ref{sec:speeds}, and~\ref{sec:offset}).
\item We develop an AIS-derived hybrid replay protocol to evaluate defense methods. The evaluation protocol uses recorded benign approaches and adds controlled direct and off-axis attack runs at selected commit ranges. Asset locations are selected before fitting, cutoffs are calibrated only on benign validation encounters, and evaluation reports false alarms, detection, and warning with uncertainty (Sections~\ref{sec:motion} and~\ref{sec:calibration}).
\item We conduct an evaluation of adversarial attacks on the camera system with hull patches constrained for printability under maritime viewing transformations, and we measure detection suppression, benign relabeling, and misses caused by the maritime viewing condition. This separates camera failures covered by a simple radar presence check from benign-label attacks that require more sophisticated defenses (Sections~\ref{sec:patch} and~\ref{sec:rq1}).
\end{enumerate}

Sections~\ref{sec:related}--\ref{sec:defense} cover related work, the threat model, the data, the attackers, and the defense; Section~\ref{sec:results} the results; and Sections~\ref{sec:discussion}--\ref{sec:dual-use} their implications, limitations, and release policy. The evaluation harness, configurations, and result summaries are released as a citable artifact \cite{artifact_dataset,artifact_software}.

\section{Background and related work}\label{sec:related}

\subsection{The uncrewed surface vessel threat}\label{sec:usv-threat}

The documented use of USVs by the Houthis in the Red Sea includes converted patrol boats, converted fishing skiffs, and purpose-built hulls. The early boats were mostly patrol boats fitted with a warhead \cite{car2017,iiss2024}. Later attacks have also used purpose-built Toufan craft and fishing vessels \cite{iiss2024,cardash2024}. The fishing vessels resemble the civilian fishing boats that sail that coast every day \cite{cardash2024}, making them visually difficult to detect. A converted patrol boat reached 45 knots with a 454\,kg warhead, and purpose-built Toufan craft are reported at 35 to 52 knots \cite{cardash2024,iiss2024}.

The approaches to a working port carry many legitimate small craft, and the defender must identify a hostile craft operating among that traffic. Small vessels are not required to report who they are via the automatic identification system; the defender therefore has to decide from available sensor data, at a distance, which vessels present a potential threat. These decisions must be taken quickly: at speeds of 40 knots, a craft covers five nautical miles in under eight minutes.

\subsection{How vessel identity is established at sea}\label{sec:identity-channels}

The automatic identification system broadcasts a vessel's name, type, and position. Larger vessels and passenger ships must carry it under SOLAS Chapter V \cite{solas2000ais}, but small craft are not required to carry it. Even when a transponder is present, the name and vessel type are entered manually and are unreliable in practice \cite{harati2007}. Falsification has been documented in previous studies \cite{ais_spoofing}. An attacker will choose to simply transmit nothing, which is common among legitimate small boats as well. Methods that screen commercial vessels against critical infrastructure using transponder history and registry records are therefore not useful for threat detection \cite{wielgosz2025}.

Radar measures range and bearing over time and provides a track \cite{radar_tbd2025}, but it does not indicate vessel type. Cameras paired with a machine-learning detector can report a predicted vessel class. Systems of this kind are fielded for maritime surveillance, both shipboard and on shore masts \cite{seaai2025,anduril_sentry} near critical assets. Human lookouts can also identify potential threats based on appearance but cannot attend to every contact at once in a busy waterway. In a simulated closed-circuit television watch, two-thirds of operators missed a salient event inside their own field of attention, with experienced personnel performing no better than novices \cite{nasholm2014}. Physical disguise can mislead both camera systems and human lookouts.

A shore-based camera with computer vision detection is a scalable channel for assigning vessel type at useful standoff, but its output depends on vessel appearance, which is controlled by the attacker and can be modified. A confirmatory check must therefore use evidence from a modality other than appearance.

\subsection{Layered coastal defense}\label{sec:layers}

Shore defense installations often layer multiple sensing channels: transponder monitoring, radar surveillance, one or more cameras with automated detection, and a procedural response. A simple cross-modal safeguard is a radar presence check, which compares detections across channels and alarms when a radar track has no corresponding camera detection. This approach covers ordinary failures such as sensor dropout or a low-contrast contact, and it also covers attacks that suppress detection, since a craft undetected by cameras still generates a radar track.

A presence check cannot cover a vessel the camera detects but labels benign, such as a USV disguised as a fishing boat: the radar track is paired with a camera detection, so nothing alarms.

\subsection{Related work}\label{sec:prior-art}

\textbf{Detection of a USV versus detection by one.} Most maritime detection benchmarks involving USVs are built for a vessel perceiving its own surroundings for collision avoidance \cite{mods2022}. Shore defense is the reverse direction, with a near-waterline aspect and an adversarial subject trying to avoid detection or be labeled benign. The counter-uncrewed-aircraft field has a defender-side camera benchmark \cite{visiodect2026}, although it varies weather and altitude rather than including adversarial targets. No corresponding defender-side maritime benchmark is available.

\textbf{Consistency between class and motion.} The closest prior work to the defense we describe is PercepGuard. It detects cases where a camera on a self-driving car has been tricked into assigning the wrong class by asking whether that class matches how the object has moved \cite{percepguard2023}. We adopt this idea but change three things. First, PercepGuard measures motion from the camera itself by tracking the object's box in the image, so an attack on the camera can also fool the check. Our defense measures motion from radar, which cannot be easily reached by an attacker. Second, PercepGuard is trained on labeled tracks of each road class. No hostile class exists to train on here since USV radar tracks are not publicly available, so we instead model the movement of real benign vessels and develop approaches to flag vessels that fall outside those patterns. Third, the attacker in PercepGuard changes what the camera sees and leaves the object's true motion alone. An attacker in our scenario must actually move like the vessel class the camera reports to avoid detection. Other related consistency check approaches from the literature assess whether a reported class fits the other objects in the scene, from image-region features or from the detector's class list \cite{sceme2020,scenebert2021} rather than using motion.

\textbf{Attacks confined to one channel.} A second line of previous work studies what happens when an attacker can reach one sensor within a fused system and shows that the fused decision can still fail even when the other sensors are correct \cite{fusion2021}. Our method applies the same principle but two things differ. Those defenses test disagreement inside model features and report the accuracy retained under attack; we compare the camera's class with the vessel's physical motion and report the resulting constraint on the attacker. Those studies also leave the object unchanged and corrupt only a sensor reading \cite{kimghosh2019,wang2022}, whereas the attacker here actually changes the craft's appearance rather than changing the sensor reading.

\textbf{Maritime and remote-sensing adversarial work.} The closest line of related work within the maritime domain tests patches and perturbations against onboard autonomy and proposes a presence-based radar, camera and transponder fusion defense \cite{saimas2023,walter2025}. A parallel line of work attacks ship detection from overhead imagery and synthetic-aperture radar \cite{shippatch2025,sar2023}. Work on general-purpose object detection supplies a benchmark of printed adversarial patches \cite{apricot2020}. A separate benchmark of digital attacks notes that inconsistent definitions of attack success impede comparison \cite{detbench2026}.

\textbf{Vessel behavior.} Modeling normal vessel motion from transponder data and flagging departures from it is a mature, non-adversarial line of work \cite{pallotta2013,track_anomaly}. Most detectors score the track alone \cite{track_anomaly}; where expected motion is conditioned on vessel type, the type comes from the same transponder message as the track, so an attacker who controls the transponder controls both sides of the comparison \cite{pallotta2013}. 

Table~\ref{tab:related} compares these lines by viewpoint, evidence used, and defense.

\begin{table*}[tbp]
\caption{Closest lines of prior work, by viewpoint, evidence used, and defense. Among the cited studies, none pairs a shore-based view of an incoming craft with a defense that uses world-coordinate motion learned from real benign traffic.}\label{tab:related}
\footnotesize
\begin{tabularx}{\textwidth}{@{}>{\raggedright\arraybackslash}p{0.205\textwidth}>{\raggedright\arraybackslash}p{0.15\textwidth}YY@{}}
\toprule
Line of work & Viewpoint & Evidence used & Defense \\
\midrule
Remote-sensing ship attacks \cite{shippatch2025,sar2023} & Overhead & Optical or SAR imagery & Retrain on filtered SAR attacks; none for optical patches \\
Onboard maritime autonomy \cite{saimas2023,walter2025} & Onboard, at sea & Camera; radar and transponder & Presence-based fusion \\
Onboard USV benchmarks \cite{mods2022} & Onboard, detection \emph{by} a craft & Camera, multisensor & None (non-adversarial) \\
Counter-uncrewed-aircraft benchmarks \cite{visiodect2026} & Aerial, detection \emph{of} a drone & Camera & None (non-adversarial) \\
Class--motion consistency \cite{percepguard2023} & Onboard, driving & Image-plane box sequences & Supervised class--track check \\
Single-source robustness \cite{kimghosh2019,fusion2021} & General, driving & Multimodal features & Cross-modal inconsistency in feature space \\
Vessel behavior modeling \cite{pallotta2013,track_anomaly} & Track level & Transponder tracks & Behavior model, not adversarial \\
\textbf{This work} & \textbf{Shore, detection \emph{of} an incoming craft} & \textbf{Camera class plus world-frame track, measured against the defended asset} & \textbf{Benign-only class--motion check; attacker must change physical motion} \\
\bottomrule
\end{tabularx}
\end{table*}

\section{Threat model and problem statement}\label{sec:threat}

The attacker has two goals against the camera. The first is suppression: to prevent the craft from being detected by the camera. The second is disguise: to allow the craft to be detected but labeled as fishing, sailing, or recreational rather than being flagged as malicious. Suppression is effective against a camera system alone, but under a layered defense it creates a mismatch between the observed radar track and camera detections. Disguise does not create this mismatch, and it is therefore the residual threat when a radar presence check is in place and is the case evaluated by the consistency checks in this paper.

The primary attack vector is physical disguise as described in Section~\ref{sec:usv-threat}: a small hull converted or finished so that shore cameras classify it as ordinary traffic. The attacker may separately choose how the craft moves, such as a slow transit followed by acceleration toward the asset or a direct run-in at burst speed. Under the threat model, radar continues to measure the resulting speed, course, and closing rate, and the attacker cannot falsify that measurement.

The method by which an attacker achieves a benign label does not affect the consistency check. A converted hull, a hull disguised as a yacht or speedboat, or a successful adversarial patch would all produce the same input: a benign camera class paired with a radar track. We therefore evaluate the defense conditional on a benign camera label without placing a patch on the hull; however, we still evaluate patches because a successful patch could change the class without requiring the structural changes needed for physical disguise.

The patch attacker is given one of three levels of access to the camera model. Under white-box access the attacker knows the model architecture and trained weights and can use this knowledge to construct the patch. Gray-box access means the attacker knows the model family (a YOLO detector in this study) and constructs the patch against a same-family model of a different size. Black-box access means cross-architecture transfer: the attacker develops the patch on one model and is scored on a separate held-out model architecture. We use within-family transfer (gray-box) as the default, white-box access as an upper bound, and cross-architecture transfer (black-box) as a separate condition.

The evaluation of the patch attacker and of the disguised-hull attacker is bounded as follows. Historical AIS is used only offline, to learn how real traffic moves. In deployment the runtime motion comes from radar, so spoofing a transponder cannot affect the check; in this evaluation AIS tracks stand in for radar throughout, a substitution whose consequences are set out in Section~\ref{sec:limitations}. All attacks and all scoring are digital: we do not place patches on real craft or run field trials. We assume that radar tracks are already formed and that each is matched to the camera detection of the same craft. What remains at the attack--defense interface is the class the camera assigns together with features of the radar track.

To evaluate the motion checks, we retain the recorded benign prefix of an AIS-derived encounter to a selected commit range and replace the remainder with a simulated attack run-in. We call the resulting hybrid track a replay (Section~\ref{sec:motion}). The scenario factors are the camera-assigned class, commit range, run-in speed, approach geometry, and report-gap treatment. Attacker access is varied only in the separate patch experiment. Table~\ref{tab:metrics} separates these controlled factors from defensive responses and operational outcomes.

\begin{table}[tbp]
\caption{Scenario factors, defensive responses, and operational outcomes used in the evaluation. Camera outcomes are scored in the adversarial-patch evaluation; detection, false alarms, and warning are scored on the consistency checks.}\label{tab:metrics}
\begin{tabularx}{\linewidth}{@{}>{\raggedright\arraybackslash}p{0.28\linewidth}Y@{}}
\toprule
Metric & Definition \\
\midrule
Scenario factors & Camera-assigned class; commit range; run-in speed; direct, constant-angle, or dog-leg geometry; and empirical or nominal report gaps. \\
Defensive responses & Presence-check outcome; speed-cutoff flag; closing-rate flag; or abstention when a class or scoring window is unavailable. \\
Evasion success rate (ESR) & Fraction of attacks that hide a previously detected craft from the camera. \\
Targeted misclassification (TMSR) & Fraction of attacks that leave the craft detected but labeled as a benign class, given that the clean image was labeled hostile. \\
Detection rate & Fraction of replays flagged at a checkpoint at or after the commit and before the craft arrives. Reported for the speed-cutoff and closing-rate checks. \\
Flag before the commit & Fraction of replays first flagged before the commit, while the track is still the recorded approach. A false alarm on real motion, not counted as detection. \\
False-alarm rate (FAR) & Fraction of real benign encounters the defense flags at any checkpoint, on the same rule and the same cutoffs. \\
Close-approach FAR & Among benign encounters that come within a given range of the asset, the fraction flagged before they first do. The benign counterpart of a flag before a commit at that range. \\
Warning after the commit & Minutes of travel remaining at the first flag at or after the commit. A miss counts as zero. Bounded by the run-in duration, which is reported with it. \\
\bottomrule
\end{tabularx}
\end{table}

\section{Evaluation setup: data and detectors}\label{sec:setup}

Two datasets are used in the evaluation: shore-view vessel images for the camera detector and historical vessel tracks for the motion models. The datasets are not paired at the observation level, and neither provides observations to the other model.

The image dataset merges four public sources with a USV class dataset we curate into one {COCO}-format \cite{coco2014} detection set of 25{,}262 images and 71{,}501 labeled vessels \cite{seaships,smd2017,aboships,mcships}. SeaShips and the on-shore videos of the Singapore Maritime Dataset are shore surveillance. ABOShips is shot from a moving vessel. McShips is collected from the web. The curated USV class contributes 261 images and 300 labeled craft, assembled from public web and video frames. The onboard images are used only for training so that evaluation includes only the shore view a port camera would see. The train/validation/test split is 22{,}804 / 1{,}229 / 1{,}229 images, with near-duplicates and video sequences kept inside a single split.

The assembled dataset has two limits that impact the results that follow. The USV and fishing classes have few labeled examples. Recreational speedboats are labeled as such only in onboard imagery, so that class has no shore-camera test: shore small boats in the Singapore set are tagged only as generic small craft.

The track dataset contains one week of MarineCadastre AIS data from the United States coast, 1--7 June 2023: 153{,}811 voyage tracks from 30{,}053 vessels, 59.1\% of them Class B \cite{marinecadastre}. The data is cleaned by removing invalid coordinates and speeds, thinned to one report a minute, and split into voyages at gaps longer than thirty minutes. Vessel class motion models and thresholds are fit on the 95{,}849 tracks from 13{,}174 benign vessels.

The camera's class names and the transponder's self-reported ship types are mapped into the canonical classes in Table~\ref{tab:classes}. The speedboat disguise evaluated below maps to \texttt{recreational}, the powered-pleasure-craft class. This mapping allows the defense methods evaluated to pair a camera-assigned class with a motion model fit on transponder-typed tracks. \texttt{small\_craft} uses a pooled Class~B proxy because AIS has no corresponding ship-type code. 

\begin{table*}[tbp]
\caption{Canonical classes shared by camera detections and AIS-derived motion models. Camera examples are labels from the merged image set. The reported attack replays claim fishing, sailing, or recreational; the remaining scoreable classes contribute benign calibration and false-alarm encounters.}\label{tab:classes}
\footnotesize
\begin{tabularx}{\textwidth}{@{}>{\raggedright\arraybackslash}p{0.20\textwidth}>{\raggedright\arraybackslash}p{0.21\textwidth}>{\raggedright\arraybackslash}p{0.28\textwidth}Y@{}}
\toprule
Canonical class & Camera examples & AIS motion-model source & Evaluation use \\
\midrule
\texttt{fishing} & Fishing boat, trawler & Type 30 (fishing) & Replay label; benign calibration and FAR \\
\texttt{sailing} & Sailboat & Type 36 (sailing) & Replay label; benign calibration and FAR \\
\texttt{recreational} & Motorboat; evaluated speedboat disguise & Type 37 (pleasure craft) & Primary replay label; benign calibration and FAR \\
\texttt{passenger\_}\allowbreak\texttt{ferry} & Passenger ship, ferry, cruise ship & Types 40--49 and 60--69 (high-speed and passenger craft) & Benign calibration and FAR \\
\texttt{cargo\_merchant} & Cargo, container, bulk, and tanker vessels & Types 70--89 (cargo and tanker) & Benign calibration and FAR \\
\texttt{working\_}\allowbreak\texttt{service} & No dedicated class in the image set & Types 31--34, 50--55, and 58 (towing, pilot, rescue, service) & Motion model retained; current detector does not assign this class \\
\texttt{small\_craft} & Generic boat, kayak, miscellaneous small boat & Class~B tracks labeled fishing, sailing, or recreational & Scoreable through a pooled proxy; not a reported replay label \\
Other detector classes & Generic civilian, military, USV, aids, and unknown objects & No benign class-conditional model & Scorer abstains \\
\bottomrule
\end{tabularx}
\end{table*}

As a proxy for proprietary commercial shore camera systems, we train representative open-source detectors using real-time single-stage and transformer model families. The results are representative of a camera layer of this kind rather than a specific fielded system type.

Three detectors are trained on the merged image set \cite{yolo11,rtdetr2024}. Two YOLO11 models (small and large) are the surrogates the patch is crafted against. A real-time transformer detector (RT-DETR, large) is held out entirely and is the black-box target in Section~\ref{sec:threat}. On undisguised USVs in the shore test set the detectors find the craft 91--96\% of the time and assign it the correct class 84--96\% of the time. We therefore evaluate the defense on a camera that can already see these craft when they are not disguised.

\section{Test scenarios and attacker models}\label{sec:attacks}

We model two types of attackers. The first applies a printability-constrained adversarial patch to the hull to hide the craft or change its class. The second uses a physically disguised hull and chooses how the craft moves toward the asset.

\subsection{Adversarial patch attacker}\label{sec:patch}

The adversarial patch is a square placed on the lower hull within the detected vessel box, with its texture kept smooth and close to a fixed set of printable colors \cite{brown2017,dpatch2018,eykholt2018,sharif2016}. The patch spans roughly 35\% of the shorter side of the target box, with a range of 20--45\%. Each optimization step averages the loss over four draws that vary the viewing conditions, patch size, and placement. Only craft whose shortest box side is at least 32 pixels are attacked.

A single optimizer is used with a different loss for suppression versus relabeling. Suppression reduces the true-class score until every prediction overlapping the craft falls below the detection threshold. Relabeling reduces the true-class score while increasing the score of a selected benign class, so a detection remains under the target label. Shapira et al.\ used this loss to change a car detection to a bus detection with a patch placed on the car; they report success in over 90\% of both digital and physical trials \cite{shapira2022}.

A shore camera records vessels under varying viewing conditions. We apply seven image transformations: scale, rotation, motion blur, glare, spray, a low camera angle, and a rough sea. During optimization the severity of each transform is randomly selected. Evaluation uses a five-step scale from none (L0) to severe (L4). The same transform implementations are used during optimization and evaluation. This is a maritime application of expectation over transformation \cite{athalye2018}.

At the high end of the viewing conditions scale, the viewing change alone can hide a craft that was detected in the clean image. A raw miss rate then misattributes the effect of weather as a successful patch. We therefore report both a raw rate as the fraction of attacks that hide a previously detected craft and a second rate that uses only those images the viewing change itself did not already hide. If the craft is already hidden before the sticker is applied, that miss is not counted as a successful patch. The miss rates in Section~\ref{sec:rq1} use the second rate.

\subsection{Replay construction}\label{sec:motion}

Detection rates and warning times in Section~\ref{sec:results} are reported for simulated attacks derived from track replays because no recorded track of a USV attack is publicly available. Each simulated attack begins with a replay of a recorded benign track followed by a simulated attack run-in. The encounter begins at the last report beyond 12\,nm before the craft enters the 6\,nm evaluation ring (Section~\ref{sec:calibration}). Recorded reports are retained until the craft first crosses a selected commit range. A source track qualifies only if it crosses that range from outside before leaving the ring or before one hour has elapsed.

At the commit range, the replay replaces the remainder of the recorded track with a straight, constant-speed run-in ending at 0.3\,nm. Position, speed, and course scatter are sampled from segments of the training tracks. Reports retain the gaps from the source track, and the first simulated report occurs one full source gap after the final recorded report. This construction keeps the recorded prefix fixed while varying class, commit range, speed, geometry, and report-gap treatment. 

\begin{figure}[tbp]
  \centering
    \includegraphics[width=\linewidth]{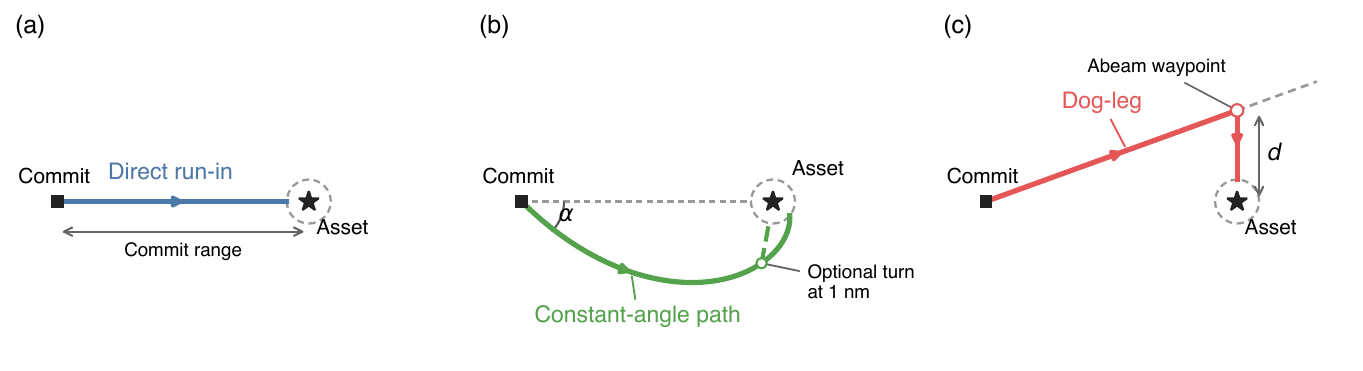}
    \caption{Replay geometry after the commit (schematic, not to scale). The star marks the protected asset, the square marks the commit, and the dashed circle marks the 0.3\,nm endpoint. (a)~A direct run-in holds the bearing to the asset. (b)~A constant-angle path is continuously re-aimed and spirals inward; the dashed segment shows the optional turn directly toward the asset at 1\,nm. (c)~A dog-leg aims at a waypoint $d=1$ or 2\,nm abeam and then turns inward; the gray dashed segment shows its continuation without that turn.}\label{fig:motion-geometry}
\end{figure}

Figure~\ref{fig:motion-geometry} summarizes the geometries of the three replay attacks. The straight run-in holds the bearing to the asset, so its closing rate is about its speed. Twelve further copies of the test set keep the recorded approach and change only the segment after the commit: eight constant-angle paths and four dog-legs. 

A constant-angle path updates its heading at every simulated step to maintain a fixed angle from the current bearing to the protected asset. Because the angle is less than 90 degrees, the craft closes on the asset while curving around it in an inward spiral. Its closing rate stays near the speed times the cosine of the angle. The tested angles are 45 and 60 degrees at 20\,kn and 60 and 70 degrees at 30\,kn. 

A dog-leg instead runs straight toward a waypoint 1 or 2\,nm abeam of the asset, perpendicular to the bearing at the commit, and then turns directly toward the asset. Each dog-leg is run at 20 and 30\,kn with the waypoint 1\,nm abeam, and again at both speeds with the waypoint 2\,nm abeam. A source track that commits at or inside the waypoint distance is omitted, because the waypoint would not lie ahead of the craft. On both shapes, heading changes are limited to 8 degrees per second, and the side of the offset is drawn at random.

The commit range determines how much of the recorded approach is retained. Commits at 0.5, 1, 2, and 4\,nm retain the recorded approach until the craft is close to the asset. At 6\,nm the simulated run-in begins at the edge of the evaluation ring, and at 12\,nm it begins at the start of the encounter.

The class-derived run-in speed is computed from AIS training trips. Each trip contributes its 95th-percentile speed over ground, and the class-derived speed is the 95th percentile of those values, so one training trip in twenty is at least that fast. Run-in speed is either the class-derived speed---15.6\,kn for fishing, 30.3\,kn for recreational, and 9.2\,kn for sailing---or a hull burst speed of 42 or 45\,kn. In the training data, 449 tracks reach a 95th-percentile speed of at least 25\,kn on a nearly straight path, and 40 reach 40\,kn. The burst speeds are those of two fielded hulls. The Magura V5 cruises at roughly 22\,kn and has a top speed of 42\,kn \cite{wertheim2025}. The converted fishing skiff's speed is unclear \cite{iiss2024}, so it is scored at the 45\,kn burst reported for a converted patrol boat \cite{cardash2024}. Sailing and fishing craft are unlikely to reach 42\,kn, so burst replays under those labels test the check under an implausible class--speed pairing. The recreational class is the plausible burst-speed case.

\section{Defense methods}\label{sec:defense}

A class--motion consistency defense evaluates whether a craft of the class the camera assigns would plausibly follow the track the radar measures. Two approaches are developed and evaluated: threshold approaches using track-only features and comparison scorers using learned benign vessel motion models. The vessel motion models are fit on the 95{,}849 benign tracks of Section~\ref{sec:setup}, and every cutoff is set on validation encounters.

\subsection{Consistency scorer}\label{sec:scorer}

For each detected vessel, the motion consistency scorer receives the camera's predicted class and a window of the vessel's track. It returns a flag or an abstention. Rather than selecting the closest class, we estimate the range of normal motion for each class from benign examples and test whether the observed motion falls within that range \cite{scholkopf2001}. This design does not require hostile training tracks and also avoids the closed-set behavior of ship-type classifiers \cite{yan2022} which assign every track to some class even when the track is outside the normal range of all classes.

The threshold checks use a per-class cutoff on one feature. The track-only speed-cutoff check flags when the 95th percentile of reported speed over the window exceeds the claimed class's cutoff. The asset-relative closing-rate check flags when the 90th percentile of closing rate toward the asset exceeds the class's cutoff. We refer to them below as the speed-cutoff and closing-rate checks. If the window is too short to compute the features, or the predicted class has no cutoff, the scorer abstains. For comparison, we fit a Gaussian mixture per class \cite{laxhammar2008}, an isolation forest \cite{isolationforest2008}, and Mahalanobis distance to each class mean \cite{mahalanobis1936,ledoitwolf2004}. These scorers use the full feature sets below, abstain on the same contacts, and follow the same calibration procedure (Section~\ref{sec:calibration}; Table~\ref{tab:scorers}).

\subsection{Track-only features}\label{sec:track-only}

The speed-cutoff check uses only the radar track and the longest complete window among 120, 180, and 300\,s. Its feature is the 95th percentile of speed over ground. The comparison scorers also use the median and standard deviation of speed; the fraction of reports below 0.5\,kn, which represents loitering; path straightness, defined as the great-circle distance from the first report to the last divided by total path length; and mean absolute acceleration in knots per minute. When course over ground is available, they also use mean turn rate in degrees per second and circular standard deviation of course.

\subsection{Asset-relative motion}\label{sec:geometry}

Features based only on track without knowledge of asset location cannot distinguish a fast approach toward the asset from a recreational craft passing nearby. The closing-rate check therefore measures motion relative to the asset's position. Its cutoff is on closing rate; the comparison scorers also use the track-only features, bearing stability, closest distance and time to closest approach, and the fraction of the window spent closing. These are standard closest-point-of-approach and constant-bearing-decreasing-range measures used to assess collision risk \cite{colregs1972,cpa_tcpa}. Threat-identification systems also score a small craft's closest approach when visual identification cannot cover every radar track \cite{istis2007}. When paired with the camera class, a closing rate that is expected for a merchant ship approaching a berth may be flagged as unusual for a craft assigned the recreational class.

These features require a specified asset location, and that location affects the false-alarm rate: a point in open water has few approaches, while a point in a busy channel has frequent benign approaches. We therefore preselect the asset positions before model fitting or false-alarm evaluation \cite{nosek2018}. The positions represent berths or anchored ships in five port regions and are computed from each port's AIS traffic. A fairway point in heavy traffic and an offshore terminal are excluded from fitting and used only as held-out stress-test locations. A track is paired with an asset only when it enters the 6\,nm ring around that asset on an inbound leg; otherwise the asset-relative features are not computed.

These features are computed on the longest complete window among 300 and 600\,s. Fishing, sailing, and cargo have too few training encounters at 300\,s and thus use 600\,s alone. The inbound leg contains every report in the window through the closest approach, which must fall inside the 6\,nm ring, and requires four reports, including three before closest approach. Reports from before ring entry count toward the leg, so a run that began outside the ring is scored on the approach that brought it inside. Because the asset-relative features are defined only for a contact approaching the asset, the closing-rate check is not triggered outside the 6\,nm ring. The speed-cutoff check has no such gate and scores from 12\,nm.

\subsection{Calibration and evaluation discipline}\label{sec:calibration}

Detection and warning time are evaluated over a sequence of windows. Each track is paired with every asset whose 6\,nm ring it enters and is scored from the last report beyond 12\,nm before entry until it leaves the ring, one hour passes, or the track ends. Checkpoints occur every 60\,s once enough history is available, and the first checkpoint exceeding the cutoff raises the flag. Each class's cutoff targets a five percent false-alarm rate on validation encounters. Classes with fewer than 50 scoreable validation encounters use a cutoff pooled across the check's encounters. Speed cutoffs are 21.9\,kn for recreational craft, 21.4\,kn for working and service craft, 29.8\,kn for passenger ferries, and a pooled 25.5\,kn for fishing, sailing, and cargo. Closing-rate cutoffs are 10.9, 10.4, and 28.0\,kn for the first three classes, 7.6\,kn for sailing, and a pooled 14.1\,kn for fishing and cargo. The comparison scorers are calibrated in the same way.

The reported checks use the class-conditional cutoffs above. To evaluate the impact of conditioning the cutoffs on class, every class is assigned one shared cutoff per check, 25.5\,kn for speed and 14.1\,kn for closing rate, using the same validation encounters and the same five percent false-alarm target. Classes with fewer than 50 scoreable validation encounters already use these pooled values. The comparison applies them to every class, including those with enough encounters to have a cutoff of their own: recreational craft, working and service craft, passenger ferries, and, for closing rate, sailing craft.

The cutoffs are set on benign validation encounters alone. No observed or simulated attack trajectory is used in calibration, and no threshold is tuned on the replays. Before the commit, each replay is identical to its recorded benign encounter, so a flag in that segment is a false alarm. 

Intervals on the sequential encounter rates are 95\% Wilson intervals with the number of distinct source trips as the sample size. Replays built from one trip share its recorded prefix across commit ranges and hulls, so they are not independent, and counting each trip once gives intervals wider than counting each replay would.

\section{Results}\label{sec:results}

\subsection{Susceptibility of camera detection to adversarial attack}\label{sec:rq1}

\begin{figure}[tbp]
  \centering
    \includegraphics[width=\linewidth]{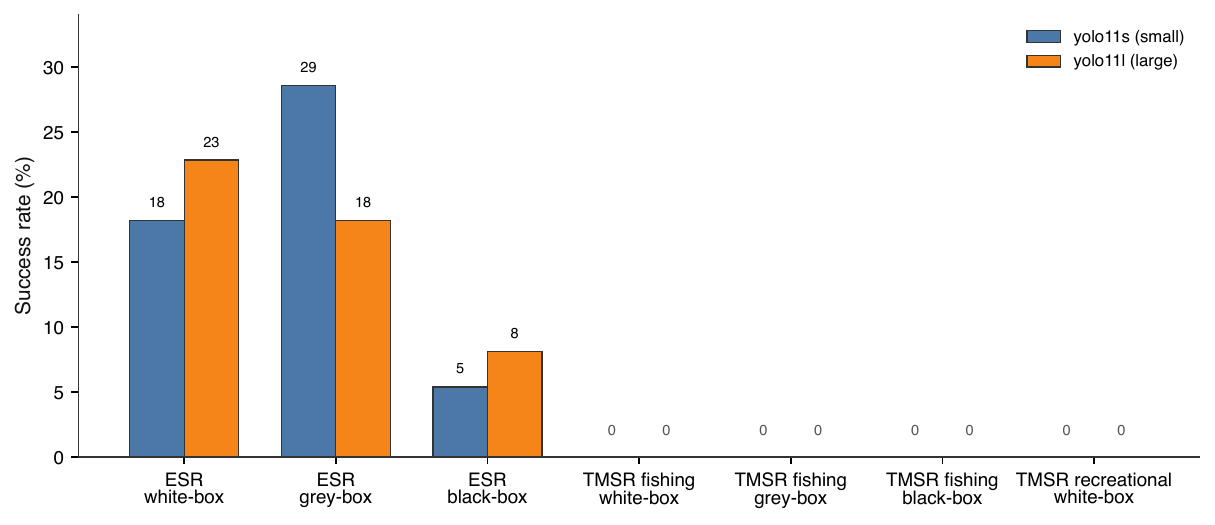}
    \caption{Suppression (evasion success rate, ESR) and hostile-to-benign relabeling (targeted misclassification rate, TMSR) at the base viewing condition, grouped by access to the camera model. Suppression occurs at every access level; no tested configuration produces a benign-label change.}\label{fig:rq1}
\end{figure}

We find that a printability-constrained hull patch suppresses some camera detections but is unable to change any of them to a benign class label. Figure~\ref{fig:rq1} reports both outcomes at the base viewing condition, grouped by the attacker's access to the camera model.

With white-box access the miss rate is 18\% on the small YOLO detector and 23\% on the large one (6 of 33 and 8 of 35). With gray-box access it is 18--29\%, and against the held-out transformer it falls to 5--8\%. Each rate is calculated over the 33--37 images that the corresponding detector already detects cleanly. The image counts are small, so the intervals are wide, and single-digit differences between access levels or detectors are not meaningful. Glare and spray increase the miss rate by up to 22 percentage points at the most severe setting, while rough sea produces little change. Under black-box access the miss rate peaks at 16.7\%, below every white-box and gray-box base-condition rate. Scale, rotation, motion blur, and low camera angle suppress some detections without a patch at high severity, so those cells measure the viewing condition rather than the patch (Section~\ref{sec:patch}).

None of the evaluated attacks succeed in creating a hostile-to-benign label change across the 420 attempts covering all access levels and both target classes. At the base condition, the detector still reports a USV on most attacked images (26--27 of 33 on the small YOLO; 26--27 of 35 on the large). Images that lose the USV box do not gain an overlapping fishing or recreational box and are counted as suppression successes.

A suppressed camera detection leaves a radar track without a matching camera detection, which the assumed presence check flags. The remainder of the results evaluates the more severe threat: a hostile craft assigned a benign camera label through physical disguise (Section~\ref{sec:threat}).

\subsection{Defending against disguise: speed and closing-rate checks}\label{sec:rq2}

Physically disguised attackers are not flagged by a simple radar presence check and so require more sophisticated defense methods for detection. The rates reported below therefore measure improvement over a layered defense including a radar presence check rather than over camera detection alone.

\begin{figure}[t]
  \centering
    \includegraphics[width=\linewidth]{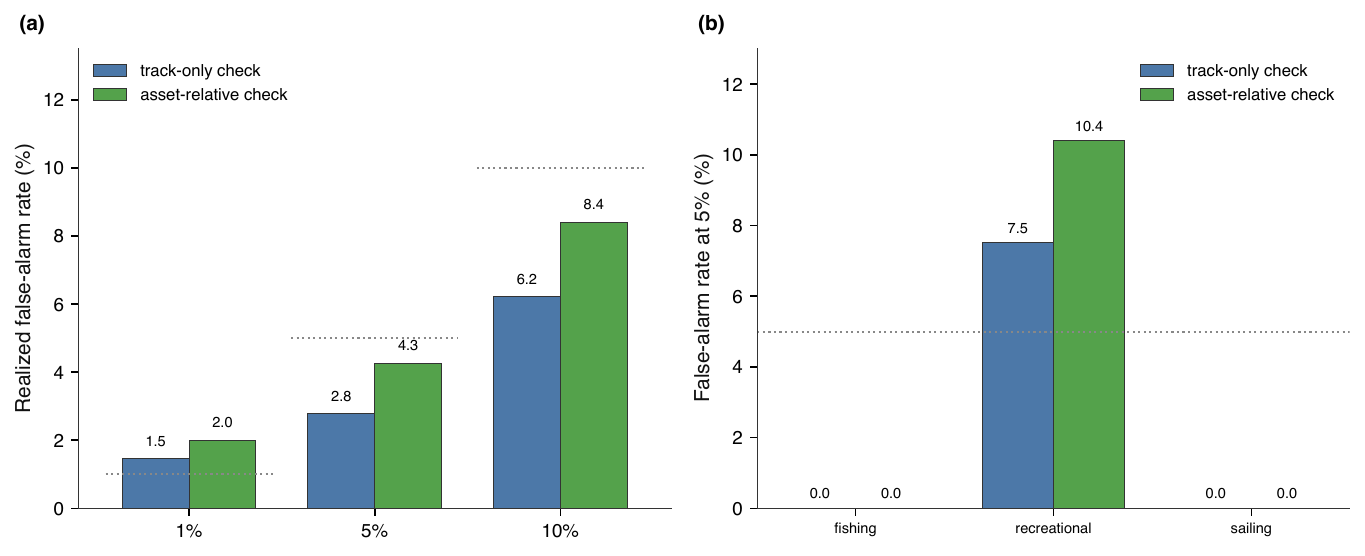}
    \caption{Any-time false-alarm rate on held-out encounters. (a)~Both checks at the 1\%, 5\%, and 10\% targets. (b)~The 5\% target by class; no fishing or sailing encounter alarms on either check.}\label{fig:far}
\end{figure}

Recreational replays account for 104 of the 121 cases at class-derived speeds that commit within 4\,nm. The speed-cutoff check flags 96\% (95\% interval 86--99\%) and the closing-rate check flags 99\% (91--100\%) with overall false-alarm rates of 2.8\% and 4.3\% (Figure~\ref{fig:far}, Table~\ref{tab:given-label-ddr}). The recreational class-derived speed is 30.3\,kn, so a run-in at that speed is above the 21.9\,kn cutoff. That cutoff is the level that 95\% of recreational validation encounters near the assets remain under across all scored windows. Fishing and sailing class-derived speeds of 15.6 and 9.2\,kn are below the pooled 25.5\,kn speed cutoff. The speed-cutoff check flags none of those replays, while the closing-rate check flags 7 of 8 fishing and 9 of 9 sailing replays. Across all three classes, median warning after the commit is 1.1 minutes for the speed-cutoff check and 2.3 minutes for the closing-rate check, on run-ins with median durations of 3.4 and 4.5 minutes.

Before the simulated run-in begins, 14\% of replays are flagged by the speed-cutoff check and 27\% by the closing-rate check. Because these flags occur in the corresponding recorded benign part of the track, they are false alarms rather than detections. Tracks eligible to serve as replay sources must contain enough reports and approach the asset closely. They therefore form a harder-than-average benign subset, which the checks flag more often than the broader benign test set. Across all benign encounters from the three replay classes, 2.8\% and 6.0\% are flagged before first reaching 4\,nm (Table~\ref{tab:given-label-ddr}).

\begin{table}[tbp]
\caption{Test replays at class-derived speeds that commit within 4\,nm, using cutoffs calibrated to the nominal 5\% validation target. Detection is a flag at or after the commit; brackets are 95\% Wilson intervals over source trips. The any-flag row also includes pre-commit false alarms. The replay-only row excludes flags that occur at the same checkpoint in the corresponding recorded benign encounter. The closing-rate check scores 119 of the 121 replays (102 of 104 recreational). Fishing and sailing replays come from three and five trips.}\label{tab:given-label-ddr}
\begin{tabularx}{\linewidth}{@{}>{\raggedright\arraybackslash}Xlll@{}}
\toprule
 & n & Speed cutoff & Closing rate \\
\midrule
Fishing & 8 & 0\% [0, 56] & 88\% [34, 99] \\
Sailing & 9 & 0\% [0, 43] & 100\% [57, 100] \\
Recreational & 104 & 96\% [86, 99] & 99\% [91, 100] \\
All replays & 121 & 83\% [71, 90] & 98\% [91, 100] \\
Any flag, including before the commit & 121 & 83\% & 99\% \\
Flag before the commit & 121 & 14\% & 27\% \\
Replay-only run-in flag & 121 & 81\% [69, 89] & 98\% [91, 100] \\
Warning after the commit & 121 & 1.1\,min & 2.3\,min \\
False-alarm rate, all encounters & --- & 2.8\% [2, 4] & 4.3\% [3, 6] \\
False-alarm rate, source encounters & --- & 16\% & 31\% \\
\bottomrule
\end{tabularx}
\end{table}

\subsection{Attackers below the speed cutoff}\label{sec:speeds}

To test run-ins below the 21.9\,kn recreational speed cutoff, we replay the same test sources and commits at 10, 15, and 20\,kn without recalibration (Table~\ref{tab:scorers}). At 15 and 20\,kn, the closing-rate check flags 94\% and 95\% of recreational-labeled run-ins that commit within 4\,nm because their closing rates exceed the 10.9\,kn cutoff. For run-ins beginning at 6 or 12\,nm, detection rises to 97--100\%, with 15 to 21 minutes of warning. The speed-cutoff check flags only 11--12\% at all three speeds, and most of those flags also occur in the corresponding recorded benign encounters and thus are considered false alarms. At 10\,kn, the closing-rate check's 27\% post-commit rate is the same kind of flag: the run-in stays below the cutoff once the window contains only simulated reports and so these flags should also be treated as false alarms.

The fitted comparison scorer models detect less at every speed. The asset-relative isolation forest is closest, at 80--81\% for 15--20\,kn; the mixture and Mahalanobis distance flag 12\% or less at those speeds with either feature set.

\begin{table}[tbp]
\caption{Scorers on the same test replays, using cutoffs calibrated to the nominal 5\% validation target. FAR is the held-out false-alarm rate. Columns 10 to 30.3\,kn report post-commit detection for recreational-labeled replays that commit within 4\,nm; 30.3\,kn is the recreational class-derived speed. The one-cutoff rows assign every class the shared validation cutoff. The either row flags a replay when either frozen check does and uses no new cutoff. All rates are reported to one decimal place.}\label{tab:scorers}
\footnotesize
\setlength{\tabcolsep}{4pt}
\begin{tabular*}{\linewidth}{@{\extracolsep{\fill}}lrrrrr@{}}
\toprule
Scorer & FAR & 10\,kn & 15\,kn & 20\,kn & 30.3\,kn \\
\midrule
Speed cutoff & 2.8\% & 11.4\% & 11.5\% & 11.5\% & 96.2\% \\
Closing-rate cutoff & 4.3\% & 27.2\% & 94.2\% & 95.1\% & 99.0\% \\
Either check & 4.9\% & 26.7\% & 93.3\% & 94.2\% & 99.0\% \\
Speed, one cutoff & 2.0\% & 10.5\% & 10.6\% & 10.6\% & 92.3\% \\
Closing rate, one cutoff & 5.1\% & 22.3\% & 91.3\% & 90.3\% & 94.1\% \\
\midrule
Mixture, track-only & 3.8\% & 1.0\% & 1.9\% & 1.9\% & 9.6\% \\
Mixture, asset-relative & 6.4\% & 7.8\% & 6.8\% & 7.8\% & 61.8\% \\
Mahalanobis, track-only & 3.0\% & 8.6\% & 7.7\% & 11.5\% & 65.4\% \\
Mahalanobis, asset-relative & 4.3\% & 9.7\% & 9.7\% & 10.7\% & 80.4\% \\
Isolation forest, track-only & 4.4\% & 10.5\% & 9.6\% & 14.4\% & 81.7\% \\
Isolation forest, asset-relative & 5.9\% & 21.4\% & 79.6\% & 80.6\% & 94.1\% \\
\bottomrule
\end{tabular*}
\end{table}

\subsection{Off-axis approaches}\label{sec:offset}

A straight run-in closes at a rate roughly equal to its speed, so the 15--20\,kn result above is the benign closing-rate quantile. An attacker who is aware of the check can instead aim off the asset. The twelve copies in Section~\ref{sec:motion} are scored here with the same cutoffs (Table~\ref{tab:offset}).

On recreational-labeled replays that commit within 4\,nm, a 20\,kn constant-angle path maintained at 60 degrees from the bearing to the asset closes at 10.1\,kn. The closing-rate check flags 28 of 103 and the speed check flags 12 of 105. From 12\,nm, where the whole encounter follows the constant-angle path, the closing-rate check flags none of 30 and the union flags 1 of 30, but the attack's length is prolonged to 70 minutes. From 6\,nm it lasts 34 minutes and the closing-rate check flags 20 of 48, but 19 of those flags fall at the commit range and so are considered false alarms. A path whose closing rate stays below 10.9\,kn takes at least $(R - 0.3)/10.9$ hours to arrive from range $R$ in nautical miles, about 31 minutes from 6\,nm. The same closing rate at 30\,kn requires a 70-degree offset and is above the 21.9\,kn speed cutoff, so the speed check flags 104 of 105 within 4\,nm. From 6\,nm it flags 47 of 48 with 31 minutes of warning on a 33-minute run-in, and from 12\,nm it flags all 30 with 65 minutes of warning on a 68-minute run-in.

Turning directly toward the asset at 1\,nm restores closing-rate detection of the 20\,kn, 60-degree path, at 96 of 103 flagged within 4\,nm, but the median warning is just 0.5 minutes, and from 12\,nm it is about 0.1 minutes. The flag is the turn, not the long constant-angle approach. On the 30\,kn, 70-degree path the speed check has already flagged the oblique portion, so the turn does not determine the warning.

A dog-leg aimed 1 or 2\,nm abeam does not lower closing rate while leaving speed high. The off-axis leg still closes at 18--29\,kn. Among recreational dog-legs that commit within 4\,nm, the closing-rate check flags 78 of 80 at 20\,kn aimed 1\,nm abeam and all 46 aimed 2\,nm abeam. At 30\,kn the speed check flags 78 of 80 and all 46, and the closing-rate check flags all of them. A modest lateral offset is not enough to stay under the cutoff.

The union of the two frozen checks flags a replay when either check does, and it uses no new cutoff. On the straight run-ins it tracks the better of the two checks (Table~\ref{tab:scorers}). Its held-out false-alarm rate is 4.9\% (37 of 757), against 2.8\% and 4.3\% for the checks alone, because 16 of the speed check's 21 alarms are also closing-rate alarms; 12.3\% of held-out recreational encounters are flagged (31 of 253). It does not flag the 20\,kn, 60-degree leg once the scoring window is that leg. Where its straight-run rate is a point below the closing-rate check, the denominator includes a replay that check does not score.

The 10\,kn closing rate sits above the sailing cutoff of 7.6\,kn and below the fishing cutoff of 14.1\,kn, so the closing-rate check flags 8 of 9 sailing replays but none of the 8 fishing replays on the same 20\,kn, 60-degree tracks. 

\begin{table}[tbp]
\caption{Off-axis recreational-labeled replays committing within 4\,nm, scored with the frozen cutoffs. Speed detection and closing detection are the percentages flagged by the corresponding check at or after the commit and before arrival; either detection is the percentage flagged by at least one check. Leg p90 is the median 90th-percentile closing rate during the off-axis leg. Warning reports median post-commit warning from either check, with misses assigned zero, followed by median run-in duration. The closing-rate check may not score every replay, so its denominator can be one below $n$.}\label{tab:offset}
\footnotesize
\begin{tabularx}{\linewidth}{@{}>{\raggedright\arraybackslash}Xrrrrrr@{}}
\toprule
Shape & Leg p90 & $n$ & \shortstack{Speed\\detection} & \shortstack{Closing\\detection} & \shortstack{Either\\detection} & Warning \\
\midrule
20\,kn, 45$^\circ$ & 14.2\,kn & 104 & 12\% & 90\% & 89\% & 5.3 of 7.1\,min \\
20\,kn, 60$^\circ$ & 10.1\,kn & 105 & 11\% & 27\% & 27\% & 0 of 10.1\,min \\
20\,kn, 60$^\circ$, turn at 1\,nm & 10.1\,kn & 104 & 12\% & 93\% & 92\% & 0.5 of 8.0\,min \\
30\,kn, 70$^\circ$ & 10.3\,kn & 105 & 99\% & 27\% & 99\% & 8.2 of 9.9\,min \\
30\,kn, 70$^\circ$, turn at 1\,nm & 10.3\,kn & 104 & 97\% & 97\% & 99\% & 5.3 of 7.2\,min \\
20\,kn dog-leg, 2\,nm abeam & 18.4\,kn & 46 & 11\% & 100\% & 100\% & 16 of 18\,min \\
30\,kn dog-leg, 2\,nm abeam & 27.5\,kn & 46 & 100\% & 100\% & 100\% & 10 of 12\,min \\
\bottomrule
\end{tabularx}
\end{table}

\subsection{Effect of the camera class}\label{sec:class-value}

The checks above give each camera class its own cutoff when the validation sample is large enough. Fishing and sailing speeds, and the fishing closing rate, use the shared cutoff because those classes have fewer than 50 validation encounters. The recreational speed cutoff (21.9\,kn), the recreational closing-rate cutoff (10.9\,kn), and the sailing closing-rate cutoff (7.6\,kn) are applied separately. 

As a comparison, we rescore the same checkpoints with a single pooled cutoff per check, 25.5\,kn and 14.1\,kn, assigned to every class (Section~\ref{sec:calibration}). Replacing the recreational closing-rate cutoff with the shared cutoff reduces detection by 2.9 percentage points at 15\,kn (95\% source-trip interval 0.0--6.3) and 4.9 points at 20\,kn (1.1--8.6; Table~\ref{tab:scorers}). Class conditioning therefore improves detection modestly but does not create the main 15--20\,kn result.

The largest observed effect of class conditioning is for sailing, although the sample is thin. Its class-derived speed of 9.2\,kn exceeds the 7.6\,kn sailing cutoff but not the shared 14.1\,kn cutoff, so detection falls from 9 of 9 replays to none. Those replays come from five source trips. Fishing is unchanged because it already uses the shared cutoff on both checks.

Shared cutoffs redistribute false alarms rather than uniformly reducing them. The shared speed cutoff lowers the overall held-out rate from 2.8\% to 2.0\%. The shared closing-rate cutoff lowers the recreational rate from 10.4\% to 8.8\%, but replacing the 28.0\,kn passenger-ferry cutoff with 14.1\,kn flags 12 of 151 ferry encounters and raises the overall rate from 4.3\% to 5.1\%.

\subsection{Disguise cost and warning time}\label{sec:rq3}

\begin{figure}[tbp]
  \centering
    \includegraphics[width=\linewidth]{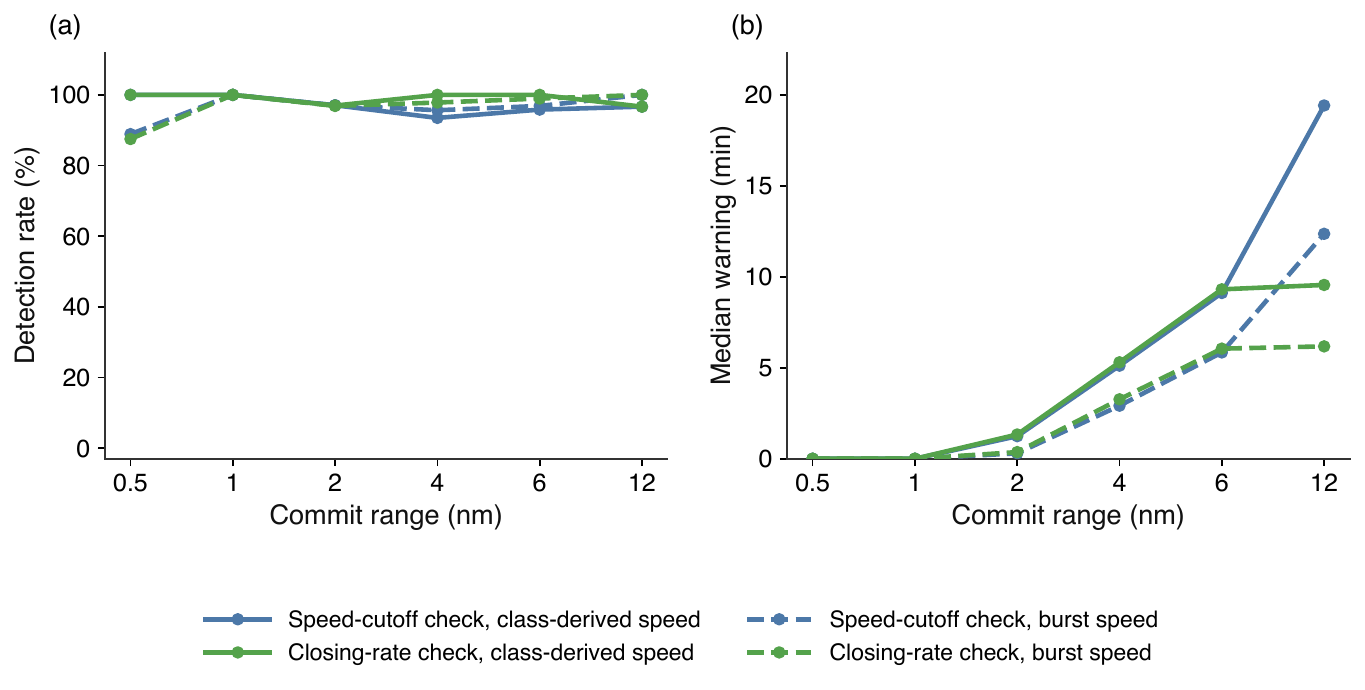}
     \caption{Recreational-labeled test replays by commit range. Commits at 6 and 12\,nm begin at the ring edge or encounter start. Solid lines use the class-derived speed (30.3\,kn), and dashed lines use a 42--45\,kn burst. (a)~Detection. (b)~Median warning after the commit, with misses assigned zero.}\label{fig:rq3}
\end{figure}

The defender's warning cannot be greater than the duration of the run-in after the commit, and so a later commit leaves less warning time. The commit range defined in Section~\ref{sec:motion} determines how long the replay retains the recorded benign approach. Table~\ref{tab:commit} and Figure~\ref{fig:rq3} report recreational-labeled replays at each commit range, at class-derived and burst speeds.

At the class-derived speed, both checks flag at least 93\% of run-ins from every commit range; moving the commit inward therefore reduces warning rather than detection (Table~\ref{tab:commit}). Commits at 0.5 or 1\,nm leave no warning before arrival. From 12\,nm, the speed-cutoff check leaves 19.4 minutes of warning, compared with 9.6 minutes for the closing-rate check, because asset-relative features are available only inside the 6\,nm ring. For burst run-ins beginning at 6 or 12\,nm, pooled over the three classes, the speed-cutoff and closing-rate checks flag 185 and 187 of 188 replays, respectively.

For the tested direct approaches, closing rate has a larger effect on detection than commit range. Detection falls sharply below the 10.9\,kn recreational cutoff (Section~\ref{sec:speeds}); crossing the 6\,nm ring at 10\,kn takes 36 minutes.

Sampling report gaps from the training distribution, whose median is about three minutes, lowers detection by 3--5 percentage points for run-ins committing within 4\,nm and by 10--14 points for run-ins beginning at 12\,nm. Median warning from 6 or 12\,nm shortens by at most 1.1 minutes. Sparser reports delay flags because a checkpoint can be triggered only when a report arrives. Higher-rate radar updates may reduce this delay, but that requires validation with radar data.

Fishing and sailing class-derived speeds are below the pooled speed cutoff, so the speed-cutoff check does not flag any of their run-ins beginning at 6 or 12\,nm. The closing-rate check flags all eight replays from each class, with median warnings of 20 and 35 minutes on run-ins lasting 33 and 56 minutes. Neither check flags any held-out fishing or sailing encounter (0 of 43 and 0 of 52). These replays come from only three and four trips, so neither rate supports a class-level conclusion.

\begin{table}[tbp]
\caption{Recreational-labeled test replays by commit range, using cutoffs calibrated to the nominal 5\% validation target. Each cell reports detection with its 95\% Wilson interval over source trips, followed by median post-commit warning where nonzero; misses are assigned zero warning. Burst replays share source trips with replays at the class-derived speed for the same commit. Run-in is the median time from commit to arrival. The closing-rate check scores one fewer replay at 0.5 and 2\,nm at the class-derived speed, and two and four fewer at burst speed.}\label{tab:commit}
\begin{tabularx}{\linewidth}{@{}lccYY@{}}
\toprule
Commit & n & Run-in & Speed cutoff & Closing rate \\
\midrule
\multicolumn{5}{@{}l}{Class-derived speed, 30.3\,kn} \\
0.5\,nm & 9 & 0.3\,min & 100\% [70, 100] & 100\% [68, 100] \\
1\,nm & 15 & 1.3\,min & 100\% [78, 100] & 100\% [78, 100] \\
2\,nm & 34 & 3.2\,min & 97\% [84, 100]; 1.2\,min & 97\% [84, 99]; 1.3\,min \\
4\,nm & 46 & 7.2\,min & 93\% [81, 98]; 5.1\,min & 100\% [91, 100]; 5.3\,min \\
6\,nm & 48 & 11\,min & 96\% [84, 99]; 9.1\,min & 100\% [91, 100]; 9.3\,min \\
12\,nm & 30 & 23\,min & 97\% [81, 99]; 19.4\,min & 97\% [81, 99]; 9.6\,min \\
\midrule
\multicolumn{5}{@{}l}{Burst, 42--45\,kn} \\
0.5\,nm & 18 & 0.2\,min & 89\% [56, 98] & 88\% [53, 98] \\
1\,nm & 30 & 0.9\,min & 100\% [78, 100] & 100\% [78, 100] \\
2\,nm & 68 & 2.2\,min & 97\% [84, 100]; 0.3\,min & 97\% [84, 99]; 0.4\,min \\
4\,nm & 92 & 4.9\,min & 96\% [84, 99]; 2.9\,min & 98\% [87, 100]; 3.3\,min \\
6\,nm & 96 & 7.6\,min & 97\% [86, 99]; 5.8\,min & 99\% [89, 100]; 6.1\,min \\
12\,nm & 60 & 16\,min & 100\% [86, 100]; 12.4\,min & 100\% [86, 100]; 6.2\,min \\
\bottomrule
\end{tabularx}
\end{table}

\section{Discussion}\label{sec:discussion}

\subsection{Principal findings and interpretation}\label{sec:disc-findings}

The camera results determine which layer of the defense must respond. The tested adversarial hull patches suppressed some detections but never produced a benign label. Suppression leaves an unmatched radar track for the presence check, whereas a benign label leaves a matched detection that must be flagged by other means. Physical disguise can create that label without attacking the model, so the absence of patch-based relabeling narrows but does not remove the threat that the motion checks address.

The class-conditional motion checks constrain how a craft can approach the protected asset rather than guarantee identification as hostile. Direct run-ins are usually caught by a closing-rate check, while a fast oblique approach can reduce closing rate but remains visible to the speed check. This complementarity is geometric: benign recreational craft may move quickly while passing an asset, producing a higher speed cutoff, while steering off-axis reduces only the asset-directed component of velocity and leaves speed visible. The simple speed and closing-rate checks outperform the fitted multivariate scorers on the tested direct run-ins.

Staying below both speed and closing-rate cutoffs requires a slower approach, while turning toward the asset late can preserve evasion until little warning remains but also requires a longer approach. The defense therefore imposes a trade-off between speed, path, and warning time rather than guaranteeing detection of every approach. 

\subsection{Simulation credibility and intended use}\label{sec:credibility}

Simulation credibility is specific to the decision the model is meant to support \cite{sargent2013,dodi500061}. The intended use here is comparative test and evaluation: determine how frozen speed and closing-rate checks respond when selected scenario factors change, and identify conditions that require a higher-fidelity test. The study is not an accreditation of the defense for deployment and does not estimate an operational probability of detecting a USV attack.

The replay construction is verified by exact agreement between each replay prefix and its recorded source encounter (Section~\ref{sec:motion}). Its inputs are empirical: prefixes, report gaps, and kinematic scatter come from recorded AIS, and every cutoff is calibrated on benign encounters (Section~\ref{sec:calibration}). Its fidelity is limited by the addition of the synthetic post-commit motion tracks and through the use of AIS as a stand-in for associated radar tracks (Section~\ref{sec:limitations}). The results therefore support prioritizing closing rate and speed for higher-fidelity defense test and evaluation rather than deployment decisions, consistent with US Department of Defense guidance that verification and validation effort be tailored to intended use and decision risk \cite{dodi500061,dodm5000102}.

\subsection{Relation to prior work}\label{sec:disc-prior}

The findings extend the prior work on class--motion and sensor-fusion defenses against adversarial attacks. PercepGuard checks a camera label against image-plane motion measured by the camera, while this work compares the label with motion from an independent sensor, learns from benign tracks, and requires the attacker to change the craft's movement to achieve consistency and avoid detection \cite{percepguard2023}. It also differs from maritime anomaly methods in which the vessel type and motion come from the same transponder \cite{pallotta2013,track_anomaly}, and reinforces that an uncompromised sensor is useful only when cross-channel disagreement is explicitly evaluated \cite{kimghosh2019,fusion2021,wang2022}.

Unlike the high targeted-relabeling rates reported by Shapira et al.\ for cars \cite{shapira2022}, the tested adversarial hull patches suppressed detections but did not produce a label change. Smaller targets, scarcer training data, and different detectors or viewing conditions may explain the difference, but these factors were not isolated in the experiment so as to attribute cause.

\subsection{Operational implications}\label{sec:disc-operations}

The proposed class-motion consistency checks should be considered as prioritization tools for alerting, not as guarantees of hostile vessel detection or as an autonomous interdiction method. Deployment of the methods requires a camera-assigned class, a correctly associated radar track, and the location of the protected asset. Cutoffs should be calibrated from local benign traffic and held fixed during operation. Performance should be reported by class as well as overall: the combined check's held-out false-alarm rate is 4.9\% across all encounters but 12.3\% on recreational traffic, the class most relevant to a fast disguised craft. That difference may determine whether an installation produces a manageable alert load.

Detection is operationally useful only when it leaves sufficient warning time to take action. Commits at 0.5 or 1\,nm provide no warning before arrival in the evaluated replays, whereas an approach that is flagged for inconsistency at the evaluation-ring edge or farther out can leave several minutes of warning time. A shore defense should therefore set thresholds based on its response plans and available response methods. 

Layered systems also need separate reports for separate failure modes. Suppression, benign relabeling, presence-check alarms, and class--motion inconsistency flags describe different points in the attack path. Collapsing them into a single attack-success rate can hide whether an attack that defeats the camera is stopped by another layer. Reporting each outcome separately makes the residual vulnerability visible.

\subsection{Future work}\label{sec:disc-future}

The first priority for future work is evaluation of the methods on synchronized shore-camera and radar data. Such data would test whether the cutoffs remain effective under radar noise, fragmentation, and camera--radar association errors. The evaluation should include small craft that do not carry AIS. Multi-season and multi-port collection is needed to determine how much local calibration is required and whether class-specific false-alarm patterns persist.

The attack evaluation should also shift from digital replays to physical trials. Field experiments could test whether a disguised or patched hull receives the intended camera label, how real acceleration and maneuvering affect warning, and whether an attacker can combine low closing rate with late turns or weaving more effectively than the evaluated paths. Synchronized maritime video would also permit a direct comparison with image-plane consistency methods such as PercepGuard.

Finally, warning time should be connected to human and procedural response. Experiments with operators could measure whether alerts at the observed ranges improve identification and intervention in practice without creating alarm fatigue. The same asset-relative principle could also be tested in non-maritime settings such as substations, airport perimeters, or pipeline terminals with thresholds and response-time measures adjusted appropriately to each setting.

\section{Limitations}\label{sec:limitations}

Every consistency-check detection rate and warning time is computed on AIS-derived track replays because no dataset of recorded USV attacks is publicly available. The pre-commit phase contains recorded benign motion, but the post-commit run-in is synthetic: speed changes instantaneously and then remains constant along one of the evaluated paths. The study does not test gradual acceleration, weaving, adaptive maneuvers, sea-state-dependent platform dynamics, or whether a disguised craft could reproduce the pre-commit benign track. Burst replays can also pair a claimed class with a hull speed that may be physically implausible. The reported rates compare the frozen checks within the tested factor space and should not be interpreted as field-performance estimates.

The replay sample is small and correlated. The 121 replays at class-derived speeds that commit within 4\,nm come from 56 source trips, and commit ranges and attack variants reuse source tracks. Recreational craft account for 104 cases; the fishing and sailing samples are too small to support class-level conclusions.

Motion is evaluated on one week of United States AIS data standing in for radar. AIS excludes non-reporting small craft and does not reproduce radar noise, detection probability, false plots, fragmentation, reacquisition, association errors, or radar update rates. The evaluation also assumes that every camera detection is matched to the correct radar track, which may be difficult in crowded approaches. A deployment would require synchronized local radar and camera data, recalibrated cutoffs, testing of track formation and association, and an operationally representative alerting workflow.

The camera evaluation is likewise narrow. The hostile class contains 261 curated images, and each attack rate uses only the few dozen images. The digital evaluation tests one patch recipe, one same-family transfer condition, and one held-out architecture. The absence of benign relabeling therefore does not establish that maritime relabeling is impossible.

Finally, false alarms are uneven across traffic classes. The combined check flags 4.9\% of held-out encounters overall but 12.3\% of recreational encounters. Because recreational craft are also the most plausible label for a fast disguised hull, the overall rate may understate the alert burden at facilities with frequent high-speed small-craft approaches.

\section{Dual use and responsible release}\label{sec:dual-use}

The reported speed, path, and warning-time trade-offs could inform attackers as well as defenders, although physical disguise is already documented in fielded attacks (Section~\ref{sec:usv-threat}). To support auditing without materially lowering the barrier to attack, the release includes the evaluation harness, configurations, and frozen result summaries but withholds trained detector weights, fitted per-class motion models, and high-fidelity patch templates. The Data availability statement lists the deposited materials.

\section{Conclusion}\label{sec:conclusion}

This study developed a hybrid replay protocol for comparative test and evaluation of class--motion consistency defenses against adversarial attacks by USVs on shore-based camera systems. The evaluation method combines recorded benign approaches with run-ins toward a protected asset to simulate hostile craft tracks. The evaluation method was used to evaluate potential defenses for disguised USVs that are not detected by a shore-based camera. 

Class-motion consistency defenses proved effective in this evaluation by forcing a trade-off between detection and constrained motion. Given a benign camera label, speed and asset-relative closing rate impose complementary constraints on an attacker. A direct approach raises a flag based on closing rate, while a fast off-axis approach leaves high speed visible. Remaining below both cutoffs can leave little to no warning but requires slowing down and extending the approach, increasing the likelihood of detection by other means. The simulation identifies which scenarios merit higher-fidelity testing, although AIS-based replays alone cannot establish operational effectiveness. Synchronized camera--radar validation, local calibration, operator-in-the-loop evaluation, and live trials are required before deployment.

\section*{Acknowledgements}
The author thanks the maintainers of the public vessel imagery datasets and of the MarineCadastre vessel traffic archive, without which the benign-behavior models in this work could not have been fit.

\subsection*{Generative AI disclosure}
During the preparation of this work, the author used Claude Opus 5 and GPT-5.6 Sol through Cursor to assist with research-software development and suggest improvements to the manuscript's clarity and organization. The author reviewed and tested the generated or modified code, verified the reported results and cited sources, edited the manuscript, and takes full responsibility for the content of the published article.

\section*{Declarations}

\subsection*{Competing interests}
The author declared no potential conflicts of interest with respect to the research, authorship, and/or publication of this article.

\subsection*{Funding}
This research did not receive any specific grant from funding agencies in the public, commercial, or not-for-profit sectors.

\subsection*{Data availability}
The evaluation harness, the configurations, the frozen result summaries, and the derived annotations and track features needed to reproduce the reported numbers are deposited and citable \cite{artifact_dataset,artifact_software}. Every source dataset remains under its original license and must be obtained from its original provider: the public image sets \cite{seaships,smd2017,aboships,mcships} and the vessel track archive \cite{marinecadastre}. The curated USV imagery is likewise not redistributed. The deposit carries its per-image provenance manifest and annotations, from which the class can be rebuilt from the original sources. Trained detector weights, the fitted per-class motion models, and the high-fidelity patch templates are withheld from the deposit for the reasons given in Section~\ref{sec:dual-use}.

\bibliographystyle{SageV}

\bibliography{main}

\begin{thebibliography}{10}
\providecommand{\url}[1]{\texttt{#1}}
\providecommand{\urlprefix}{URL }
\expandafter\ifx\csname urlstyle\endcsname\relax
  \providecommand{\doi}[1]{DOI:\discretionary{}{}{}#1}\else
  \providecommand{\doi}{DOI:\discretionary{}{}{}\begingroup
  \urlstyle{rm}\Url}\fi
\providecommand{\eprint}[2][]{\url{#2}}

\bibitem{raveendran2025}
Raveendran J.
\newblock Sea {{Denial}}: {{The Ukrainian Case Study}} and the {{Future}} of
  {{Naval Warfare}}.
\newblock \emph{Journal of Strategic Security} 2025; 18(4).
\newblock \doi{10.5038/1944-0472.18.4.2564}.

\bibitem{wertheim2025}
Wertheim E.
\newblock Ukraine's {Magura} naval drones: {Black Sea} equalizers.
\newblock \emph{US Naval Institute Proceedings} 2025; 151(9).
\newblock
  \urlprefix\url{https://www.usni.org/magazines/proceedings/2025/september/ukraines-magura-naval-drones-black-sea-equalizers}.

\bibitem{ambrey2024}
{Ambrey}.
\newblock Threat circular: First confirmed {RC-WBIED} impact since {November}
  2023.
\newblock Ambrey, 2024.
\newblock
  \urlprefix\url{https://ambrey.com/app/uploads/2024/06/AMBREY-THREAT-CIRCULAR_FIRST-CONFIRMED-IMPACT-OF-RC-WBIED-SINCE-NOVEMBER-2023.pdf}.

\bibitem{cardash2024}
Cardash M.
\newblock Houthi {USV} {WBIED} attacks targeting merchant ships in the {Red
  Sea}, {Yemen}.
\newblock \emph{Counter-IED Report} 2024; : 54--71
  \urlprefix\url{https://counteriedreport.com/wp-content/uploads/2024/11/54-71-Michael-Cardash-Houthi-USV-WBIED-Attacks-Targeting-Merchant-Ships-in-the-Red-Sea-COUNTER-IED-REPORT-Autumn-2024.pdf}.
\newblock Autumn 2024.

\bibitem{boretti2026}
Boretti A.
\newblock Unmanned surface vehicles for naval warfare and maritime security.
\newblock \emph{The Journal of Defense Modeling and Simulation: Applications,
  Methodology, Technology} 2026; 23(2): 317--324.
\newblock \doi{10.1177/15485129241283056}.

\bibitem{seaai2025}
{SEAAI}.
\newblock {AI} object detection solutions for shipping.
\newblock Vendor technical documentation,
  https://www.sea.ai/commercial/shipping/, 2026.
\newblock Accessed 5 September 2026. Optical and thermal machine vision
  marketed for small craft without {AIS}.

\bibitem{anduril_sentry}
{Anduril Industries}.
\newblock Sentry: Autonomous awareness.
\newblock Vendor technical documentation, https://www.anduril.com/sentry/,
  2026.
\newblock Accessed 6 September 2026. Shore-based mast with radar and
  electro-optical/infrared cameras; onboard machine-learning detection and
  classification of surface vessels.

\bibitem{eykholt2018}
Eykholt K, Evtimov I, Fernandes E et~al.
\newblock Robust physical-world attacks on deep learning visual classification.
\newblock In \emph{IEEE/CVF Conference on Computer Vision and Pattern
  Recognition (CVPR)}. pp. 1625--1634.
\newblock \doi{10.1109/CVPR.2018.00175}.

\bibitem{drp2021}
Sato T, Shen J, Wang N et~al.
\newblock Dirty road can attack: Security of deep learning based automated lane
  centering under {Physical-World} attack.
\newblock In \emph{30th USENIX Security Symposium (USENIX Security 21)}. USENIX
  Association.
\newblock ISBN 978-1-939133-24-3, pp. 3309--3326.
\newblock
  \urlprefix\url{https://www.usenix.org/conference/usenixsecurity21/presentation/sato}.

\bibitem{pa_survey2024}
Wei H, Tang H, Jia X et~al.
\newblock Physical adversarial attack meets computer vision: A decade survey.
\newblock \emph{IEEE Transactions on Pattern Analysis and Machine Intelligence}
  2024; 46(12): 9797--9817.

\bibitem{martinez2026}
Martinez AL, Champagne LE and LaCasse PM.
\newblock Simulating autonomous drone behaviors in an anti-access area denial
  ({A2AD}) environment.
\newblock \emph{The Journal of Defense Modeling and Simulation: Applications,
  Methodology, Technology} 2026; 23(2): 325--334.
\newblock \doi{10.1177/15485129241288236}.

\bibitem{harris2013}
Harris SP, Dixon DS, Dunn DL et~al.
\newblock Simulation modeling for maritime port security.
\newblock \emph{The Journal of Defense Modeling and Simulation: Applications,
  Methodology, Technology} 2013; 10(2): 193--201.
\newblock \doi{10.1177/1548512912439952}.

\bibitem{artifact_dataset}
Reifschneider J.
\newblock Counter-{USV} systems adversarial robustness dataset \& results,
  2026.
\newblock \doi{10.5281/zenodo.23345287}.
\newblock \urlprefix\url{https://doi.org/10.5281/zenodo.23345287}.
\newblock [dataset].

\bibitem{artifact_software}
Reifschneider J.
\newblock Counter-{USV} adversarial robustness, 2026.
\newblock \doi{10.5281/zenodo.23345946}.
\newblock \urlprefix\url{https://doi.org/10.5281/zenodo.23345946}.
\newblock [software].

\bibitem{car2017}
{Conflict Armament Research}.
\newblock Anatomy of a drone boat.
\newblock Conflict Armament Research, 2017.
\newblock
  \urlprefix\url{https://www.conflictarm.com/perspectives/anatomy-of-a-drone-boat/}.

\bibitem{iiss2024}
Paes WC, Beales E, Hinz F et~al.
\newblock Navigating troubled waters: The {Houthis}' campaign in the {Red Sea}
  and the {Gulf of Aden}.
\newblock The International Institute for Strategic Studies, {IISS} Research
  Paper, London, 2024.
\newblock
  \urlprefix\url{https://www.iiss.org/research-paper/2024/12/navigating-troubled-waters-the-houthis-campaign-in-the-red-sea-and-the-gulf-of-aden/}.

\bibitem{solas2000ais}
{International Maritime Organization}.
\newblock International convention for the safety of life at sea ({SOLAS}),
  1974, as amended --- chapter {V}, regulation 19: Carriage requirements for
  shipborne navigational systems and equipment.
\newblock International Maritime Organization, London, 2000.
\newblock Revised chapter {V} adopted December 2000; entered into force 1 July
  2002. {AIS} carriage requirements at Reg. 19.2.4.

\bibitem{harati2007}
Harati-Mokhtari A, Wall A, Brooks P et~al.
\newblock Automatic identification system ({AIS}): Data reliability and human
  error implications.
\newblock \emph{Journal of Navigation} 2007; 60(3): 373--389.
\newblock \doi{10.1017/S0373463307004298}.

\bibitem{ais_spoofing}
Androjna A, Pavi{\'c} I, Gucma L et~al.
\newblock {AIS} data manipulation in the illicit global oil trade.
\newblock \emph{Journal of Marine Science and Engineering} 2024; 12(1): 6.
\newblock \doi{10.3390/jmse12010006}.

\bibitem{wielgosz2025}
Wielgosz M and Ma{\l}yszko M.
\newblock A method for early identification of vessels potentially threatening
  critical maritime infrastructure.
\newblock \emph{Applied Sciences} 2025; 15(15): 8716.
\newblock \doi{10.3390/app15158716}.

\bibitem{radar_tbd2025}
Herrmann L, Garc{\'i}a-Fern{\'a}ndez {\'A}F, Brekke EF et~al.
\newblock A scalable hybrid track-before-detect tracking system: Application to
  coastal maritime radar surveillance.
\newblock arXiv preprint arXiv:2508.16169, 2025.
\newblock \doi{10.48550/arXiv.2508.16169}.

\bibitem{nasholm2014}
N{\"a}sholm E, Rohlfing S and Sauer JD.
\newblock Pirate stealth or inattentional blindness? {The} effects of target
  relevance and sustained attention on security monitoring for experienced and
  na{\"i}ve operators.
\newblock \emph{PLoS ONE} 2014; 9(1): e86157.
\newblock \doi{10.1371/journal.pone.0086157}.

\bibitem{mods2022}
Bovcon B, Muhovi{\v c} J, Vranac D et~al.
\newblock {MODS} --- a {USV}-oriented object detection and obstacle
  segmentation benchmark.
\newblock \emph{IEEE Transactions on Intelligent Transportation Systems} 2022;
  23(8): 13403--13418.
\newblock \doi{10.1109/TITS.2021.3124192}.

\bibitem{visiodect2026}
Ajakwe SO, Ihekoronye VU, Mohtasin G et~al.
\newblock {VisioDECT}: A robust dataset for aerial and scenario based
  multi-drone detection, identification, and neutralization.
\newblock \emph{Data in Brief} 2026; 65: 112448.
\newblock \doi{10.1016/j.dib.2026.112448}.
\newblock [dataset].

\bibitem{percepguard2023}
Man Y, Muller R, Li M et~al.
\newblock That person moves like a car: Misclassification attack detection for
  autonomous systems using spatiotemporal consistency.
\newblock In \emph{32nd USENIX Security Symposium (USENIX Security 23)}. pp.
  6929--6946.

\bibitem{sceme2020}
Li S, Zhu S, Paul S et~al.
\newblock Connecting the dots: Detecting adversarial perturbations using
  context inconsistency.
\newblock In \emph{European Conference on Computer Vision (ECCV)},
  \emph{Lecture Notes in Computer Science}, volume 12368. pp. 396--413.
\newblock \doi{10.1007/978-3-030-58592-1_24}.

\bibitem{scenebert2021}
Yin M, Li S, Cai Z et~al.
\newblock Exploiting multi-object relationships for detecting adversarial
  attacks in complex scenes.
\newblock In \emph{IEEE/CVF International Conference on Computer Vision
  (ICCV)}. pp. 7838--7847.
\newblock \doi{10.1109/ICCV48922.2021.00776}.

\bibitem{fusion2021}
Yang K, Lin WY, Barman M et~al.
\newblock Defending multimodal fusion models against single-source adversaries.
\newblock In \emph{IEEE/CVF Conference on Computer Vision and Pattern
  Recognition (CVPR)}. pp. 3340--3349.
\newblock \doi{10.1109/CVPR46437.2021.00335}.

\bibitem{kimghosh2019}
Kim T and Ghosh J.
\newblock On single source robustness in deep fusion models.
\newblock In \emph{Advances in Neural Information Processing Systems 32
  (NeurIPS)}. pp. 4815--4826.
\newblock ArXiv:1906.04691.

\bibitem{wang2022}
Wang S, Wu T, Chakrabarti A et~al.
\newblock Adversarial robustness of deep sensor fusion models.
\newblock In \emph{IEEE/CVF Winter Conference on Applications of Computer
  Vision (WACV)}. pp. 1371--1380.
\newblock \doi{10.1109/WACV51458.2022.00144}.

\bibitem{saimas2023}
Walter MJ, Barrett A, Walker DJ et~al.
\newblock Adversarial {AI} testcases for maritime autonomous systems.
\newblock \emph{AI, Computer Science and Robotics Technology} 2023; 2.
\newblock \doi{10.5772/acrt.15}.

\bibitem{walter2025}
Walter MJ, Barrett A and Tam K.
\newblock Preventing adversarial {AI} attacks against autonomous situational
  awareness: A maritime case study.
\newblock arXiv:2505.21609, 2025.

\bibitem{shippatch2025}
Liu C, Ding P, Zheng Z et~al.
\newblock Adversarial patch attack for ship detection via localized
  augmentation.
\newblock arXiv:2508.21472, 2025.

\bibitem{sar2023}
Gao W, Liu Y, Zeng Y et~al.
\newblock {SAR} image ship target detection adversarial attack and defence
  generalization research.
\newblock \emph{Sensors} 2023; 23(4): 2266.
\newblock \doi{10.3390/s23042266}.

\bibitem{apricot2020}
Braunegg A, Chakraborty A, Krumdick M et~al.
\newblock {APRICOT}: A dataset of physical adversarial attacks on object
  detection.
\newblock In \emph{European Conference on Computer Vision (ECCV)},
  \emph{Lecture Notes in Computer Science}, volume 12366. pp. 35--50.
\newblock \doi{10.1007/978-3-030-58589-1_3}.
\newblock [dataset]. Preprint arXiv:1912.08166.

\bibitem{detbench2026}
Winter A, Martini JV, Audigier R et~al.
\newblock Benchmarking adversarial robustness and adversarial training
  strategies for object detection.
\newblock arXiv:2602.16494, 2026.

\bibitem{pallotta2013}
Pallotta G, Vespe M and Bryan K.
\newblock Vessel pattern knowledge discovery from {AIS} data: A framework for
  anomaly detection and route prediction.
\newblock \emph{Entropy} 2013; 15(6): 2218--2245.
\newblock \doi{10.3390/e15062218}.

\bibitem{track_anomaly}
Wolsing K, Roepert L, Bauer J et~al.
\newblock Anomaly detection in maritime {AIS} tracks: A review of recent
  approaches.
\newblock \emph{Journal of Marine Science and Engineering} 2022; 10(1): 112.
\newblock \doi{10.3390/jmse10010112}.

\bibitem{coco2014}
Lin TY, Maire M, Belongie S et~al.
\newblock Microsoft {COCO}: Common objects in context.
\newblock In \emph{European Conference on Computer Vision (ECCV)},
  \emph{Lecture Notes in Computer Science}, volume 8693. pp. 740--755.
\newblock \doi{10.1007/978-3-319-10602-1_48}.

\bibitem{seaships}
Shao Z, Wu W, Wang Z et~al.
\newblock {SeaShips}: A large-scale precisely annotated dataset for ship
  detection.
\newblock \emph{IEEE Transactions on Multimedia} 2018; 20(10): 2593--2604.
\newblock \doi{10.1109/TMM.2018.2865686}.
\newblock [dataset].

\bibitem{smd2017}
Prasad DK, Rajan D, Rachmawati L et~al.
\newblock Video processing from electro-optical sensors for object detection
  and tracking in a maritime environment: A survey.
\newblock \emph{IEEE Transactions on Intelligent Transportation Systems} 2017;
  18(8): 1993--2016.
\newblock \doi{10.1109/TITS.2016.2634580}.
\newblock [dataset].

\bibitem{aboships}
Iancu B, Soloviev V, Zelioli L et~al.
\newblock {ABOships} --- an inshore and offshore maritime vessel detection
  dataset with precise annotations.
\newblock \emph{Remote Sensing} 2021; 13(5): 988.
\newblock \doi{10.3390/rs13050988}.
\newblock [dataset]. Data deposit: Zenodo 4736931.

\bibitem{mcships}
Zheng Y and Zhang S.
\newblock {McShips}: A large-scale ship dataset for detection and fine-grained
  categorization in the wild.
\newblock In \emph{IEEE International Conference on Multimedia and Expo
  (ICME)}. pp. 1--6.
\newblock \doi{10.1109/ICME46284.2020.9102907}.
\newblock [dataset].

\bibitem{marinecadastre}
{NOAA Office for Coastal Management}.
\newblock Nationwide automatic identification system 2023.
\newblock Marine Cadastre, 2023.
\newblock \urlprefix\url{https://marinecadastre.gov/ais/}.
\newblock [dataset]. {US} nationwide broadcast points, 1--7 June 2023.

\bibitem{yolo11}
Jocher G and Qiu J.
\newblock {Ultralytics} {YOLO11}, version 11.0.0.
\newblock Software release, https://github.com/ultralytics/ultralytics, 2024.
\newblock [software]. License {AGPL}-3.0.

\bibitem{rtdetr2024}
Zhao Y, Lv W, Xu S et~al.
\newblock {DETRs} beat {YOLOs} on real-time object detection.
\newblock In \emph{IEEE/CVF Conference on Computer Vision and Pattern
  Recognition (CVPR)}. pp. 16965--16974.
\newblock \doi{10.1109/CVPR52733.2024.01605}.

\bibitem{brown2017}
Brown TB, Man{\'e} D, Roy A et~al.
\newblock Adversarial patch.
\newblock arXiv:1712.09665, 2017.

\bibitem{dpatch2018}
Liu X, Yang H, Liu Z et~al.
\newblock {DPatch}: An adversarial patch attack on object detectors.
\newblock In \emph{AAAI Workshop on Artificial Intelligence Safety (SafeAI)},
  \emph{CEUR Workshop Proceedings}, volume 2301. pp. 74--81.
\newblock \urlprefix\url{https://ceur-ws.org/Vol-2301/paper_5.pdf}.

\bibitem{sharif2016}
Sharif M, Bhagavatula S, Bauer L et~al.
\newblock Accessorize to a crime: Real and stealthy attacks on state-of-the-art
  face recognition.
\newblock In \emph{ACM Conference on Computer and Communications Security
  (CCS)}. pp. 1528--1540.
\newblock \doi{10.1145/2976749.2978392}.

\bibitem{shapira2022}
Shapira A, Bitton R, Avraham D et~al.
\newblock Attacking object detector using a universal targeted label-switch
  patch.
\newblock arXiv:2211.08859, 2022.
\newblock \doi{10.48550/arXiv.2211.08859}.

\bibitem{athalye2018}
Athalye A, Engstrom L, Ilyas A et~al.
\newblock Synthesizing robust adversarial examples.
\newblock In \emph{International Conference on Machine Learning (ICML)},
  \emph{Proceedings of Machine Learning Research}, volume~80. pp. 284--293.

\bibitem{scholkopf2001}
Sch{\"o}lkopf B, Platt JC, Shawe-Taylor J et~al.
\newblock Estimating the support of a high-dimensional distribution.
\newblock \emph{Neural Computation} 2001; 13(7): 1443--1471.
\newblock \doi{10.1162/089976601750264965}.

\bibitem{yan2022}
Yan Z, Song X, Zhong H et~al.
\newblock Ship classification and anomaly detection based on spaceborne {AIS}
  data considering behavior characteristics.
\newblock \emph{Sensors} 2022; 22(20): 7713.
\newblock \doi{10.3390/s22207713}.

\bibitem{laxhammar2008}
Laxhammar R.
\newblock Anomaly detection for sea surveillance.
\newblock In \emph{11th International Conference on Information Fusion
  (FUSION)}. pp. 1--8.

\bibitem{isolationforest2008}
Liu FT, Ting KM and Zhou ZH.
\newblock Isolation forest.
\newblock In \emph{Eighth IEEE International Conference on Data Mining (ICDM)}.
  pp. 413--422.
\newblock \doi{10.1109/ICDM.2008.17}.

\bibitem{mahalanobis1936}
Mahalanobis PC.
\newblock On the generalized distance in statistics.
\newblock \emph{Proceedings of the National Institute of Sciences of India}
  1936; 2(1): 49--55.

\bibitem{ledoitwolf2004}
Ledoit O and Wolf M.
\newblock A well-conditioned estimator for large-dimensional covariance
  matrices.
\newblock \emph{Journal of Multivariate Analysis} 2004; 88(2): 365--411.
\newblock \doi{10.1016/S0047-259X(03)00096-4}.

\bibitem{colregs1972}
{International Maritime Organization}.
\newblock Convention on the international regulations for preventing collisions
  at sea, 1972 ({COLREGS}), as amended.
\newblock International Maritime Organization, London; 1050 {UNTS} 16, Reg. No.
  15824, 1972.
\newblock Adopted 20 October 1972; entered into force 15 July 1977.

\bibitem{cpa_tcpa}
Sang LZ, Yan XP, Wall A et~al.
\newblock {CPA} calculation method based on {AIS} position prediction.
\newblock \emph{Journal of Navigation} 2016; 69(6): 1409--1426.
\newblock \doi{10.1017/S0373463316000229}.

\bibitem{istis2007}
Stottler R, Ball B and Richards R.
\newblock Intelligent surface threat identification system ({ISTIS}).
\newblock In \emph{IEEE Aerospace Conference}. pp. 1--12.
\newblock \doi{10.1109/AERO.2007.353055}.
\newblock {IEEEAC} paper 1283.

\bibitem{nosek2018}
Nosek BA, Ebersole CR, DeHaven AC et~al.
\newblock The preregistration revolution.
\newblock \emph{Proceedings of the National Academy of Sciences} 2018; 115(11):
  2600--2606.
\newblock \doi{10.1073/pnas.1708274114}.

\bibitem{sargent2013}
Sargent RG.
\newblock Verification and validation of simulation models.
\newblock \emph{Journal of Simulation} 2013; 7(1): 12--24.
\newblock \doi{10.1057/jos.2012.20}.

\bibitem{dodi500061}
{US Department of Defense}.
\newblock {DoD} modeling and simulation verification, validation, and
  accreditation.
\newblock Department of Defense Instruction 5000.61, Office of the Under
  Secretary of Defense for Research and Engineering, 2024.
\newblock
  \urlprefix\url{https://www.esd.whs.mil/Portals/54/Documents/DD/issuances/dodi/500061p.pdf}.

\bibitem{dodm5000102}
{US Department of Defense}.
\newblock Modeling and simulation verification, validation, and accreditation
  for operational test and evaluation and live fire test and evaluation.
\newblock Department of Defense Manual 5000.102, Office of the Director,
  Operational Test and Evaluation, 2024.
\newblock
  \urlprefix\url{https://www.esd.whs.mil/Portals/54/Documents/DD/issuances/dodm/5000102m.PDF}.

\end{thebibliography}

\end{document}